\documentclass[12pt]{article}
\usepackage{graphicx}
\usepackage{hyperref}

\def \b{{\cal B}}

\def \bea{\begin{eqnarray}}
\def \beq{\begin{equation}}
\def \ca{{\cal A}}
\def \eea{\end{eqnarray}}
\def \eeq{\end{equation}}
\def \ko{K^0}
\def \kos{K^{*0}}

\def \ok{\overline{K}^0}
\def \oks{\overline{K}^{*0}}

\def \stt{\sqrt{2/3}}

\def \thet{\theta_\eta}

\def \c{\circ}

\begin{document}

\rightline{EFI 12-3}
\rightline{arXiv:1203.6014v3}
\rightline{July 2012}

\bigskip
\centerline{\bf FLAVOR-SU(3) TESTS FROM $D^0\to K^0 K^-\pi^+$}
\centerline{\bf AND $D^0\to\ok K^+\pi^-$ DALITZ PLOTS}
\bigskip
\centerline{Bhubanjyoti Bhattacharya and Jonathan L. Rosner}
\centerline{\it Enrico Fermi Institute and Department of Physics}
\centerline{\it University of Chicago, 5640 S. Ellis Avenue, Chicago, IL 60637}

\begin{quote}
The processes $D^0\to K^0 K^-\pi^+$ and $D^0\to\ok K^+\pi^-$ involve
intermediate vector resonances whose amplitudes and phases are related to each
other via flavor-SU(3) symmetry.  Dalitz plots for these two processes can
shed light on the usefulness of this symmetry in studying charm decays.
Until this year the only available data on this process came from a conference
report in 2002 by the BaBar Collaboration, but now an independent data sample
of higher statistics has become available from the CLEO Collaboration. The goal
is to predict Dalitz plot amplitudes and phases assuming flavor-SU(3) symmetry
and compare them with experiment.

An SU(3) fit can account for the relative magnitudes of the amplitudes for the
decays $D^0 \to K^{*-}K^+$ and $D^0 \to K^{*+}K^-$, but neither the current
BaBar sample (based on an integrated luminosity of 22 fb$^{-1}$) nor the CLEO
analysis has significant evidence for the decays $D^0\to K^{*0}\ok$ and
$D^0 \to \oks K^0$.  At this level one is unable to compare magnitudes and
phases with theoretical predictions.  The purpose of this Letter is to
advocate an analysis using the full BaBar sample (more than 20 times the 2002
value).  It should definitively determine whether predicted magnitudes and
phases agree with experiment.  A similar analysis should be possible with
an even larger sample of events collected by the Belle Collaboration at KEK-B.
\end{quote}

\leftline{PACS numbers:13.25.Ft, 11.30.Hv, 14.40.Lb}
\bigskip

An important contribution to the decay processes $D^0\to 3P$, where $P$
represents a pseudoscalar meson, involves the intermediate step in which the
$D$ meson first decays into a $P$ and a vector meson ($V$). The vector meson
then decays into two pseudoscalars.  In general, in a decay with three final
$P$ states the combination of any pair of final pseudoscalars may result from
the decay of a $V$ as long as charge, isospin, strangeness, etc.\ are
conserved.  Evidence of formation of such resonances is seen in Dalitz plots as
bands of events corresponding to the invariant mass-squared of the pair of
final state $P$ mesons.  As such, they provide information about the amplitude
and phase for the process $D\to P V$. Overlapping vector resonance bands on
Dalitz plots interfere according to their relative phases.

Amplitudes and phases of $D\to P V$ decays were studied in detail using
flavor-SU(3) symmetry in Ref.\ \cite{Bhattacharya:2008ke}. Relative phase
relations based on this symmetry were exploited in Refs.\ \cite{Bhattacharya:%
2010id, Bhattacharya:2010ji, Bhattacharya:2010tg} to observe its successes in
predicting interferences on several $D\to 3P$ Dalitz plots. In the present
Letter we consider the Dalitz plots for $D^0\to K^0 K^-\pi^+$ and $D^0\to \ok
K^+\pi^-$.  We predict amplitudes and phases for the relevant $D \to PV$
intermediate processes using flavor-SU(3) symmetry.  Data from the BaBar
\cite{Aubert:2002yc} and CLEO \cite{Insler:2012pm} Collaborations do not
provide strong enough evidence for the processes $D^0 \to K^{*0} \ok$
and $D^0 \to \oks \ko$ to permit a comparison of phases with predictions,
but BaBar's total data, more than twenty times the reported sample, should
be able to provide a definitive test.  The Belle Collaboration at KEK-B should
have at least as many events as the full BaBar sample.

\begin{figure}
\includegraphics[width=0.48\textwidth]{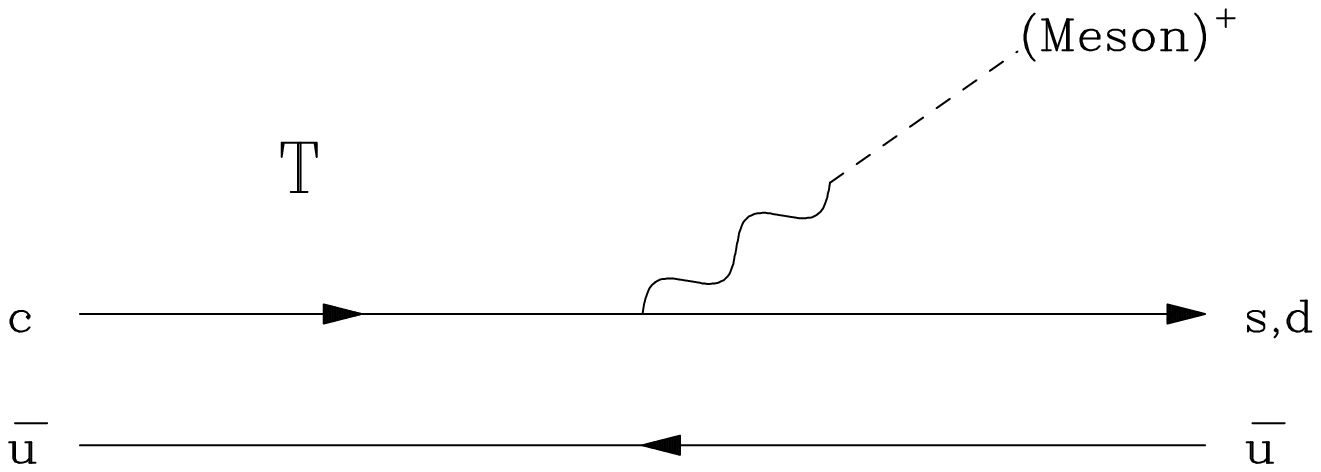}
\includegraphics[width=0.48\textwidth]{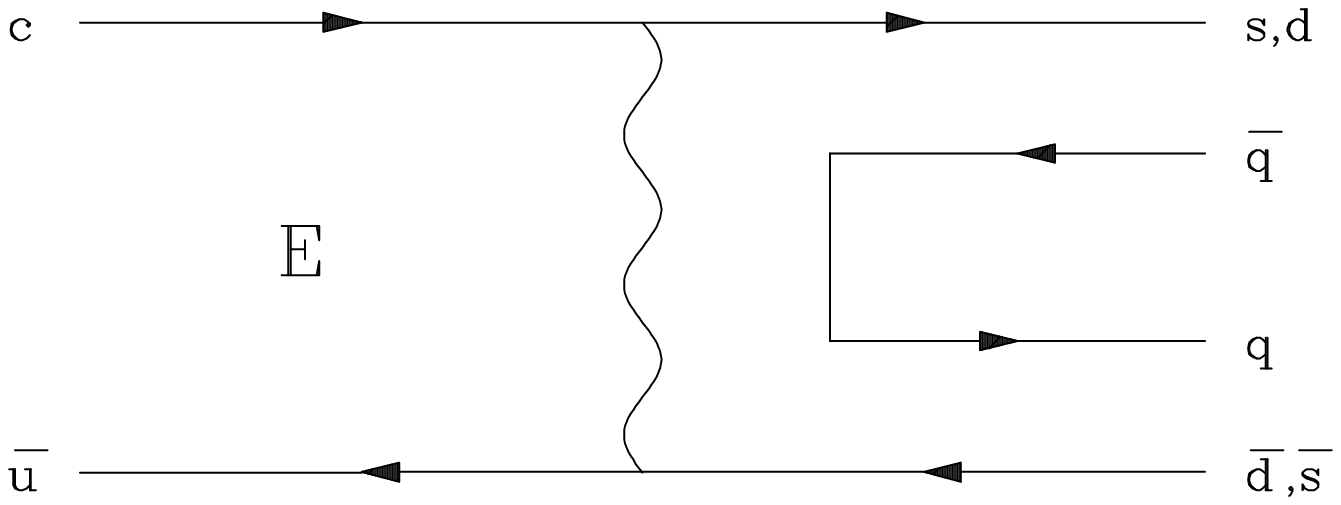}
\caption{Graphs describing tree ($T$) and exchange ($E$) amplitudes
\label{fig:TE}}
\end{figure}

We first review the flavor-SU(3) symmetry technique, and then predict
amplitudes and phases for the relevant $D \to PV$ processes, comparing them
with data.  The flavor symmetry approach used here was discussed in detail in
\cite{Bhattacharya:2008ke}.  We denote the relevant Cabibbo-favored (CF)
amplitudes, proportional to the product $V_{ud}V^*_{cs}$ of
Cabibbo-Kobayashi-Maskawa (CKM) factors, by amplitudes labeled as $T$
(``tree'') and $E$ (``exchange''), illustrated in Fig. \ref{fig:TE}.  The
singly-Cabibbo-suppressed (SCS) amplitudes, proportional to the product
$V_{us} V^*_{cs}$ or $V_{ud}V^*_{cd}$, are then obtained by using the ratio
SCS/CF $= \tan\theta_C \equiv \lambda = 0.2305$ \cite{Nakamura:2010}, with
$\theta_C$ the Cabibbo angle and signs governed by the relevant CKM factors.
The subscript $P$ or $V$ on an amplitude denotes the meson ($P$ or $V$)
containing the spectator quark in the $PV$ final state. The partial width
$\Gamma(H \to PV)$ for the decay of a heavy meson $H$ is given in terms of an
invariant amplitude $\ca$ as:
\beq
\Gamma(H \to PV) = \frac{p^{*3}}{8\pi M^2_H}|\ca|^2
\eeq
where $p^*$ is the center-of-mass (c.m.) 3-momentum of each final particle, and
$M_H$ is the mass of the decaying heavy meson. With this definition the
amplitudes $\ca$ are dimensionless.

The amplitudes $T_V$ and $E_P$ were obtained from fits to rates of CF $D \to
PV$ decays not involving $\eta$ or $\eta'$ \cite{Bhattacharya:2008ke}. To
specify the amplitudes $T_P$ and $E_V$, however, one needs information on the
$\eta - \eta'$ mixing angle ($\thet$). Table \ref{tab:tveptpev} summarizes
these results for two values $\thet = 19.5^\circ$ and $11.7^\circ$.

\begin{table}[h]
\caption{Solutions for $T_V$, $E_P$, $T_P$ and $E_V$ amplitudes in
Cabibbo-favored charmed meson decays to $PV$ final states, for $\eta$--$\eta'$
mixing angles of $\thet = 19.5^\c$ and $11.7^\circ$.
\label{tab:tveptpev}}
\begin{center}
\begin{tabular}{c c c c c} \hline \hline
 & \multicolumn{2}{c}{$\thet=19.5^\c$} & \multicolumn{2}{c}{$\thet=11.7^\c$} \\
$PV$ &  Magnitude  &  Relative  &  Magnitude  &  Relative \\
ampl.& ($10^{-6}$) & strong phase & ($10^{-6}$) & strong phase \\ \hline
$T_V$& 3.95$\pm$0.07 & Assumed 0 & \multicolumn{2}{c}{These results are}\\
$E_P$& 2.94$\pm$0.09 & $\delta_{E_PT_V} = (-93\pm3)^\circ$ & \multicolumn{2}{c}
{independent of $\thet$}\\ \hline
$T_P$ & 7.46$\pm$0.21 & Assumed 0 & 7.69$\pm$0.21 & Assumed 0 \\
$E_V$ & 2.37$\pm$0.19 &$\delta_{E_VT_V} =(-110 \pm 4)^\c$ & 1.11$\pm$0.22 &
 $\delta_{E_VT_V} =(-130 \pm 10)^\c$ \\ \hline\hline
\end{tabular}
\end{center}
\end{table}

In Tables \ref{tab:amps19} and \ref{tab:amps11} we list the $D^0 \to PV$
amplitudes relevant in Dalitz plots of interest for $\thet = 19.5^\circ$ and
$\thet = 11.7^\circ$, respectively.  Also included are their representations.
We predict the magnitudes and phases for the above amplitudes using flavor SU(3)
and compare the magnitudes with data obtained from Dalitz plot fits.

\begin{table}
\caption{Amplitudes for $D^0 \to PV$ decays of interest for the present
discussion (in units of $10^{-6}$). Here we have taken $\thet=19.5^\circ$.
\label{tab:amps19}}
\begin{center}
\begin{tabular}{c c c c c c c} \hline \hline
Dalitz & $D^0$ final & Amplitude & \multicolumn{4}{c}{Amplitude $A$} \\
 plot  & state  & representation & Re & Im & $|A|$ & Phase ($^\circ$) \\ \hline
$D^0 \to K^0 K^- \pi^+$&$K^{*+} K^-$ & $\lambda(T_P+E_V)$ & 1.533 & --0.513
 & 1.616 & --18.5 \\
 &$\oks K^0$ & $\lambda(E_V-E_P)$ & --0.151&  0.163 & 0.223 & 132.8 \\ \hline
$D^0 \to \ok K^+ \pi^-$&$K^{*-} K^+$ & $\lambda(T_V+E_P)$ & 0.875 & --0.677
 & 1.106 & --37.7 \\
 &$K^{*0} \ok$ & $\lambda(E_P-E_V)$ & 0.151 &--0.163 & 0.223 &--47.2 \\ \hline
\hline
\end{tabular}
\end{center}
\end{table}

\begin{table}
\caption{Amplitudes for $D^0 \to PV$ decays of interest for the present
discussion (in units of $10^{-6}$). Here we have taken $\thet=11.7^\circ$.
\label{tab:amps11}}
\begin{center}
\begin{tabular}{c c c c c c c} \hline \hline
Dalitz & $D^0$ final & Amplitude & \multicolumn{4}{c}{Amplitude $A$} \\
 plot  & state  & representation & Re & Im & $|A|$ & Phase ($^\circ$) \\ \hline
$D^0 \to K^0 K^- \pi^+$&$K^{*+} K^-$ & $\lambda(T_P+E_V)$ & 1.608 & --0.196
 & 1.620 & -- 6.9 \\
 &$\oks K^0$ & $\lambda(E_V-E_P)$ & --0.129& 0.481 & 0.498 &  105.0 \\ \hline
$D^0 \to \ok K^+ \pi^-$&$K^{*-} K^+$ & $\lambda(T_V+E_P)$ & 0.875 & --0.677
 & 1.106 & --37.7 \\
 &$K^{*0} \ok$ & $\lambda(E_P-E_V)$ & 0.129 &--0.481 & 0.498 &--75.0 \\ \hline
\hline
\end{tabular}
\end{center}
\end{table}

The ratio of the amplitude $|\ca(D^0 \to K^{*-}K^+)|$ relative to
$|\ca(D^0 \to K^{*+} K^-)|$ is predicted to be equal to a corresponding ratio
of Cabibbo-favored amplitudes (taken from Ref.\ \cite{Bhattacharya:2008ke}):
\beq \label{eqn:predr}
\frac{|\ca(D^0 \to K^{*-}K^+)|}{|\ca(D^0 \to K^{*+}K^-)|} =
\frac{|\ca(D^0 \to K^{*-}\pi^+)|}{|\ca(D^0 \to \rho^{*+}K^-)|}=0.685\pm0.032~.
\eeq
These ratios are less than one because the $T$ amplitudes in the numerators
involve the coupling of the weak current to a pseudoscalar meson, whose
decay constant is less than that for the vector meson involved in the
denominators:  $|T_V| < |T_P|$ (see Table I).

Flavor SU(3) predicts equal magnitudes for the much smaller amplitudes $\ca(D^0
\to \oks K^0)$ and $\ca(D^0 \to K^{*0} \ok)$:
\beq
\frac{|\ca(D^0 \to \oks K^0)|}{|\ca(D^0 \to K^{*+}K^-)|} =
\frac{|\ca(D^0 \to \ok K^{*0})|}{|\ca(D^0 \to K^{*+}K^-)|} = \left\{
\begin{array}{c} 0.138 \pm 0.033~(\thet=19.5^\circ) \cr
                 0.307 \pm 0.035~(\thet=11.7^\circ) \end{array} \right. ~.
\eeq
The predicted magnitude of these amplitudes is very sensitive to the mixing
angle $\thet$, as a result of cancellation between the amplitudes $E_V$
and $E_P$ (see Table I).

In order to obtain amplitudes from Dalitz plot fit fractions to compare with
predictions, one must recognize that the $D\to PV$ process is an intermediate
to the complete 3 body decay $D \to 3P$. The Dalitz plot fit fractions also
contain information about the vector meson decay and this must be factored out
for comparison with flavor-SU(3) predictions.  The fraction of a vector meson's
decay amplitude to a pair of $P$ mesons is given by the relevant isospin
Clebsch-Gordan factor.

\begin{table}
\caption{Conventions for the order of two pseudoscalar mesons in vector meson
decay and associated Clebsch-Gordan factors assuming the cyclic convention
of Ref.\ \cite{Insler:2012pm}
\label{tab:conv}}
\begin{center}
\begin{tabular}{c c c c c c c} \hline \hline
Dalitz Plot & \multicolumn{2}{c}{Bachelor particle} & \multicolumn{3}{c}{Vector meson decay} & $p^*$\\
&Meson&Index&Process&Indices&Clebsch factor & (in MeV)\\ \hline
&$K^0$&1&$\oks\to K^-\pi^+$&23& -- $\stt$ & 605\\
$D^0\to K^0 K^-\pi^+$&$K^-$&2&$K^{*+}\to K^0 \pi^+$&13& -- $\stt$ & 610 \\
&$\pi^+$&3&--&--&--&--\\ \hline
&$\ok$&1&$K^{*0}\to K^+\pi^-$&23& $\stt$&605 \\
$D^0\to \ok K^+\pi^-$&$K^+$&2&$K^{*-}\to\ok \pi^-$&13&   $\stt$&610 \\
&$\pi^-$&3&--&--&--&-- \\ \hline \hline
\end{tabular}
\end{center}
\end{table}

To obtain the correct Clebsch-Gordan factor including its sign, one notes that
the spin part of the amplitude for the process $D \to RC \to ABC$ ($R$
represents the intermediate resonance while $A$, $B$ and $C$ are the final
pseudoscalar mesons) is proportional to the product $\vec{p}_A\cdot\vec{p}_C$
($\vec{p}_i$ is the 3-momentum of the final state particle $i$ in the rest
frame of $R$). Since the particles $A$ and $B$ have equal and opposite
3-momenta in the resonance rest frame, this implies that swapping $A$ and $B$
while calculating the amplitude would result in an additional phase difference
of $\pi$. It is thus important to know the phase convention used to obtain the
amplitudes. In the present case, we assume a
convention employed by the CLEO Collaboration \cite{Insler:2012pm}. This
convention is presented in Table \ref{tab:conv}. Using this convention one may
then calculate the appropriate isospin Clebsch-Gordan coefficients, also noted
in Table \ref{tab:conv}.

The phase space factors for the two $D\to PV$ processes from each Dalitz plot
are not the same since the mesons involved have slightly different masses.
This very small difference, noted in Table \ref{tab:conv}, has been neglected.

The fit fractions obtained by the BaBar and CLEO analyses for relevant
intermediate $D^0\to PV$ decays corresponding to each Dalitz plot are quoted in
Table \ref{tab:data}.  We use the best CLEO fits which include the channels
$\oks \ko$ and $\kos \ok$.  Fits not including these channels actually
are superior in quality; the fit fractions for $K^{*-} K^+$ and $K^{*+} K^-$
do not differ much from those quoted.

\begin{table}
\caption{Dalitz plot fits to data from the BaBar \cite{Aubert:2002yc}
and CLEO \cite{Insler:2012pm} Collaborations
\label{tab:data}}
\begin{center}
\begin{tabular}{c c c c} \hline \hline
Dalitz Plot & $D^0$ final & \multicolumn{2}{c}{Fit fraction (\%)} \\
            &    state    & BaBar & CLEO \\ \hline
$D^0 \to \ko K^-\pi^+$ & $K^{*+} K^-$ & $63.6 \pm 5.1 \pm 2.6$ &
$67.6 \pm 6.4 \pm 3.8$ \\
 & $\oks \ko$ & $0.8 \pm 0.5 \pm 0.1$ & $1.8 \pm 1.7 \pm 0.8$ \\
$D^0 \to \ok K^+\pi^-$ & $K^{*-} K^+$ & $35.6 \pm 7.7 \pm 2.3$ &
$20.4 \pm 2.1 \pm 0.8$ \\
 & $\kos \ok$ & $2.8 \pm 1.4 \pm 0.5$ & $3.9 \pm 1.5 \pm 0.4$ \\ \hline \hline
\end{tabular}
\end{center}
\end{table}

Fit fractions quoted in Table \ref{tab:data} are normalized so as to represent
percentage of each decay mode in the specific Dalitz plots. This normalization
is different for the two different Dalitz plots. In order to compare amplitudes
for $D\to PV$ processes from two different Dalitz plots it is useful to choose
a universal normalization. To achieve this we make use of the branching
fractions for the $D\to 3P$ processes for each Dalitz plot, so as to calculate
the fraction of each $D\to PV$ process relative to a common rate or amplitude.
We thus utilize ratios of branching fractions of $D^0 \to K_S K^+ \pi^-$ and
$D^0 \to K_S K^- \pi^+$ given in Table \ref{tab:compr}.  The BaBar value has
been extracted by us from the ratios \cite{Aubert:2002yc}
\beq
\frac{\b(D^0 \to \ok K^+  \pi^-)}
     {\b(D^0 \to \ok \pi^+\pi^-)} = (5.68 \pm 0.25 \pm 0.41)\%~,~~
\frac{\b(D^0 \to \ko K^-  \pi^+)}
     {\b(D^0 \to \ok \pi^+\pi^-)} = (8.32 \pm 0.29 \pm 0.56)\%~,~~
\eeq
while the CLEO value is quoted directly by them.

\begin{table}
\caption{Comparison of ratios ${\cal B}(D^0 \to
K_S K^+ \pi^-)/{\cal B}(D^0 \to K_S K^- \pi^+)$.
\label{tab:compr}}
\begin{center}
\begin{tabular}{c c} \hline \hline
BaBar \cite{Aubert:2002yc} & CLEO \cite{Insler:2012pm} \\ \hline
$0.683 \pm 0.078$ & $0.592 \pm 0.048$ \\ \hline \hline
\end{tabular}
\end{center}
\end{table}

We make use of the data quoted in Table \ref{tab:data} and the ratios in
Table \ref{tab:compr} to calculate the relative amplitudes of the relevant
$D\to PV$ decays. The magnitudes of the amplitudes are obtained relative to
that of the process $D^0\to K^{*+} K^-$ with maximum amplitude.  These results
are listed in Table \ref{tab:compamp}. In Table \ref{tab:compamp} we also list
the predictions of magnitudes of corresponding amplitudes obtained using
flavor-SU(3) symmetry.

\begin{table}
\caption{Comparison of ratios of $D^0$ decay amplitudes extracted from Dalitz
plot fits with theoretical predictions of flavor SU(3).
\label{tab:compamp}}
\begin{center}
\begin{tabular}{c c c c c} \hline \hline
Ratio & \multicolumn{2}{c}{Experiment} & \multicolumn{2}{c}{Theory} \\
 & BaBar & CLEO & $\thet = 19.5^\circ$ & $\thet = 11.7^\circ$ \\ \hline
$\frac{|\ca(K^{*-}K^+)|}{|\ca(K^{*+}K^-)|}$ & 0.618$\pm$0.083 &
 0.423$\pm$0.037 & 0.685$\pm$0.032 & 0.685$\pm$0.032 \\
$\frac{|\ca(\oks K^0)|}{|\ca(K^{*+}K^-)|}$  & $0.159^{+0.045}_{-0.064}$ &
 $0.231^{+0.121}_{-0.231}$ & 0.138$\pm$0.033 & 0.307$\pm$0.035 \\
$\frac{|\ca(K^{*0}\ok)|}{|\ca(K^{*+}K^-)|}$ & $0.245^{+0.061}_{-0.079}$ &
 $0.261^{+0.051}_{-0.061}$ & 0.138$\pm$0.033 & 0.307$\pm$0.035 \\ \hline\hline
\end{tabular}
\end{center}
\end{table}

The success of the theoretical predictions is mixed.  While the observed ratios
quoted in Table \ref{tab:compr} and the first line of Table \ref{tab:compamp}
are less than one as predicted, the CLEO value is significantly below that of
BaBar and the predicted value (\ref{eqn:predr}).  The second and third ratios
in Table \ref{tab:compamp} are indeed seen to be small, but the evidence for
them is scant, with CLEO favoring fits without such amplitudes.

Until significant evidence for the decays $D^0 \to \oks \ko$ and $D^0 \to \kos
\ok$ is found, it is premature to compare the phases predicted in Tables II and
III with experiment.  BaBar's total sample is more than 20 times as large as
reported in Ref.\ \cite{Aubert:2002yc}, and an updated analysis would provide
much more convincing statistics.  The Belle Collaboration should have at its
disposal at least as many events as the full BaBar sample.

We thank Brian Meadows and Guy Wilkinson for helpful discussions. This work was
supported in part by the United States Department of Energy through
Grant No.\ DE FG02 90ER40560.

\end{document}